\documentstyle[sprocl,epsfig]{article}

\newcommand{\hdick}{\noalign{\hrule height1.4pt}}
\newcommand{\MeV}{{\rm MeV}}
\newcommand{\GeV}{{\rm GeV}}

\newcommand{\fb}{{\rm fb}}

\begin{document}

\title{ Determination of sparticle masses and SUSY parameters\footnote{
    Talk given at 
      LCWS~99, Sitges, Spain, 28 April -- 5 May 1999}
  }

\author{ Hans-Ulrich Martyn }

\address{ I. Physikalisches Institut, RWTH Aachen, Germany }

\author{ Grahame A. Blair }

\address{ Royal Holloway and Bedford New College,
  University of London, UK }

\maketitle
\abstracts{
  A case study  will be presented
  to determine the particle masses and parameters of a specific
  mSUGRA model at the {\sc Tesla} Linear Collider  
  with high precision.
  }
  
\section{Introduction}

\begin{figure}[hbt]
\centering
\epsfig{figure=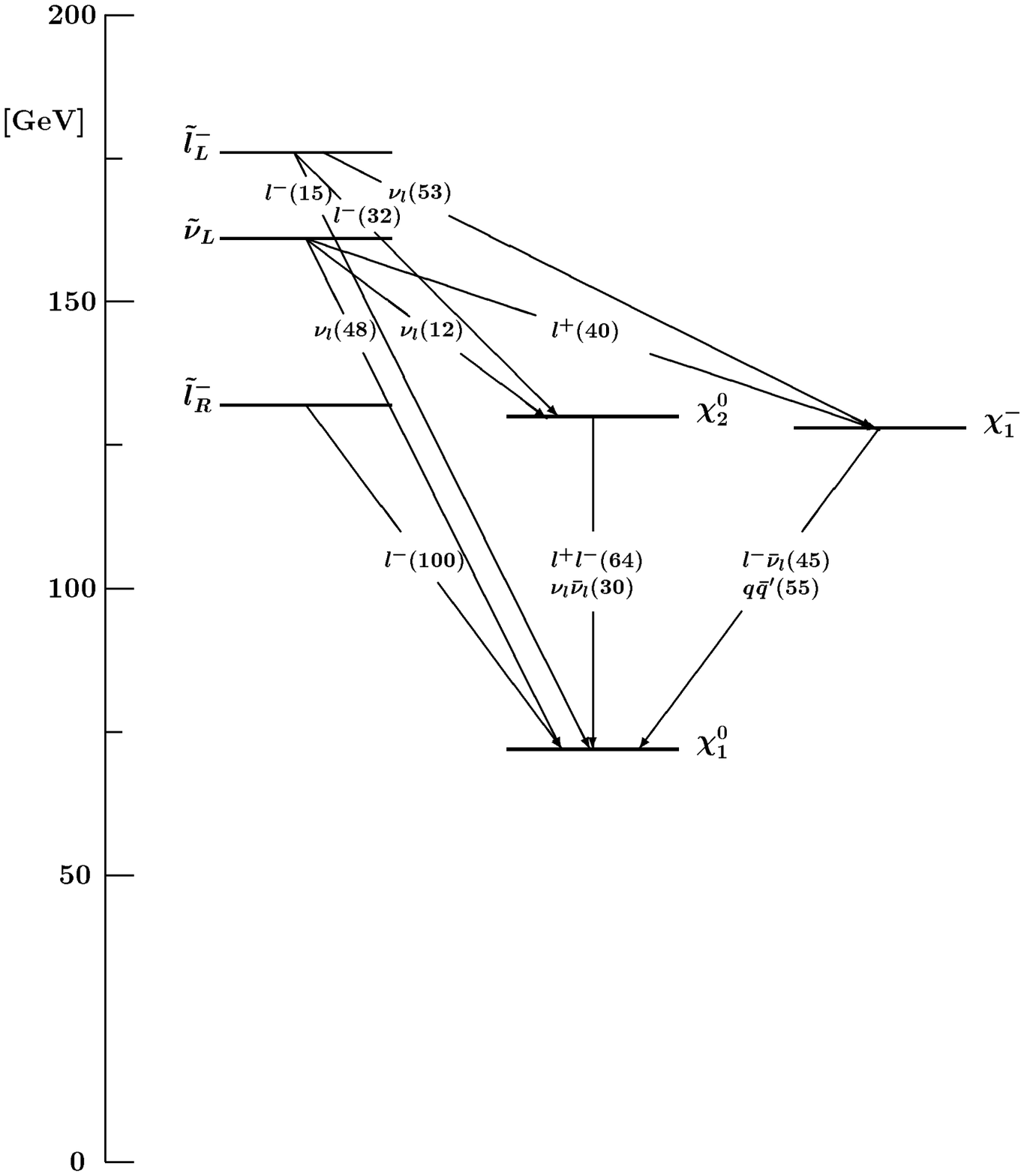,%
  bbllx=40pt,bblly=390pt,bburx=570pt,bbury=730pt,clip=,%
  height=5cm}
\vspace{-.2cm}
\caption{Mass spectrum and decay modes of sleptons and light gauginos
\label{fig:spectrum}}
\end{figure}

If supersymmetry will be realized in nature precise mass measurements of the 
particle spectrum will be very important in order to determine the underlying 
theory.
The potential of the proposed {\sc Tesla} Linear Collider~\cite{cdr} will be 
studied within a $R$-parity conserving mSUGRA scenario with parameters
$  m_0 = 100~\GeV, \ m_{1/2} = 200~\GeV, \ A_0 = 0~\GeV, \
  \tan\beta = 3$ and ${\rm sgn}(\mu) > 0 $.
In general SUSY signatures are fairly easy to detect~\cite{lcphysics}.
However, for the particle spectrum shown in fig.~\ref{fig:spectrum}
the sleptons and light gauginos will be produced simultaneously
at energies $\sqrt{s} > 360~\GeV$ and thus constitute the dominant background.
The key to disentangle the different particles and their decays is a
bottom-up approach with proper choices of the beam energies and polarisations.

Events are generated with {\sc Pythia}~6.115~\cite{pythia} 
-- which also provides masses, branching ratios and cross sections --
including QED radiation and beamstrahlung~\cite{circe}.
Polarized cross sections are calculated with {\sc Isajet}~\cite{isajet}.
The simulation is based on the detector concept of the
{\sc Tesla} CDR 
and the analyses are similar to those in ref.~\cite{cdr}.
One expects integrated luminosities of 
$500\, (320)~{\rm fb}^{-1}$ per 1--2 years 
at energies of 
$\sqrt{s} = 500\, (320)~\GeV$.
It is assumed that both beams are polarized,
 ${\cal P}_{e^-} = 0.8$ and  ${\cal P}_{e^+} = 0.6$, and the luminosity is
equally shared between $e^-_L e^+_R$ and  $e^-_R e^+_L$ runs.

\section{Mass determinations}

SUSY particles are produced in pairs and decay either directly or via cascades
into the stable neutralino $\chi_1^0$ (LSP). 
Typical signatures are multi-lepton
and/or multi-jet final states with large missing energy.
The kinematics of the decay chain allow to identify and to reconstruct the
masses of the primary and secondary sparticles.

\begin{figure}[htb]
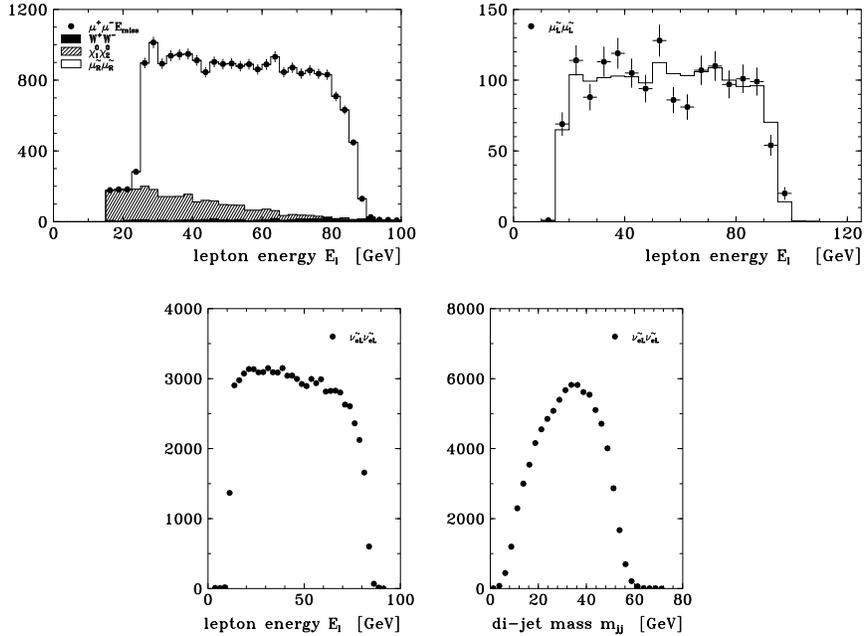

  \centering
  \epsfig{file=mur132.emu.w320.eps,%
    bbllx=10pt,bblly=0pt,bburx=470pt,bbury=750pt,clip=,%
    angle=90,width=6cm}
  \epsfig{file=mul176.emu.w500.eps,%
    bbllx=10pt,bblly=0pt,bburx=470pt,bbury=750pt,clip=,%
    angle=90,width=6cm}
  \vspace{-.3cm}
  \epsfig{file=nuel161.w500.eps,%
    bbllx=0pt,bblly=0pt,bburx=500pt,bbury=750pt,clip=,%
    angle=90,height=5cm}
  \vspace{-.2cm}
  \caption{Examples of slepton production. 
    Lepton energy spectra of
    $\tilde{\mu}_R \to \mu \chi^0_1$ at $320~\GeV$ (upper left),
    $\tilde{\mu}_L \to \mu \chi^0_2$ at $500~\GeV$ (upper right) and
    $\tilde{\nu}_{e} \to e^\mp \chi^\pm_1$ at $500~\GeV$ (lower left).
    Di-jet mass spectrum of  
    $\chi^\pm_1 \to q \bar{q}' \chi^0_1$ (lower right).}
  \label{fig:sleptons}
\end{figure}

\bigskip\noindent
{\bf Production of sleptons} \quad
The simplest case is 
$e^-_R e^+_L \to \tilde{\mu}_R \tilde{\mu}_R
             \to \mu^- \chi^0_1  \mu^+ \chi^0_1$
at $\sqrt{s} = 320~\GeV$ 
with a small background from $\chi^0_2 \chi^0_1$ production.
The energy spectrum of the decay muons is flat, see fig.~\ref{fig:sleptons},
and the end points can be 
related to the masses $m_{\tilde{\mu}_R}$ and $m_{\chi^0_1}$
with an accuracy of $\sim0.3\%$.
The partner $\tilde{\mu}_L$ can be identified at $\sqrt{s}=500~\GeV$
via a unique $6 l^\pm$ signature: 
$e^-_L e^+_R \to \tilde{\mu}_L \tilde{\mu}_L
             \to \mu^- \chi^0_2  \mu^+ \chi^0_2$
followed by $ \chi^0_2 \to l^+l^- \chi^0_1$.
Despite the low cross section of 4~fb 
such a measurement is feasible at {\sc Tesla}, giving
the masses $m_{\tilde{\mu}_L}$ and $m_{\chi^0_2}$ with
a precision of $\sim0.2\%$, see fig.~\ref{fig:sleptons}.
A final example is sneutrino production, where the flavour is tagged via
its charged decay, e.g.
$e^-_L e^+_R \to \tilde{\nu}_e \tilde{\nu}_e
             \to e^- \chi^+_1  e^+ \chi^-_1$.
The subsequent decays
$\chi^\pm_1 \to  l^\pm \nu_l  \chi^0_1$ and $ q \bar{q}' \chi^0_1$
lead to a clean $3 l^\pm + 2 j$ topology and can be used to determine
the chargino-neutralino mass difference very accurately within $0.05~\GeV$
(fig.~\ref{fig:sleptons}).
The visible cross section is huge, 
$\sigma_{LR}(\tilde{\nu}_e \tilde{\nu}_e)\,{\cal B}= 320~\fb$
compared to 
$\sigma_{LR}(\tilde{\nu}_\mu \tilde{\nu}_\mu)\,{\cal B}= 5~\fb$,
giving mass resolutions of 
$\delta m_{\tilde{\nu}_e} = \delta m_{\chi^\pm_1} = 0.1~\GeV$.

\begin{figure}[htb]
  \centering
  \mbox{ \hspace{-.3cm}
  \epsfig{file=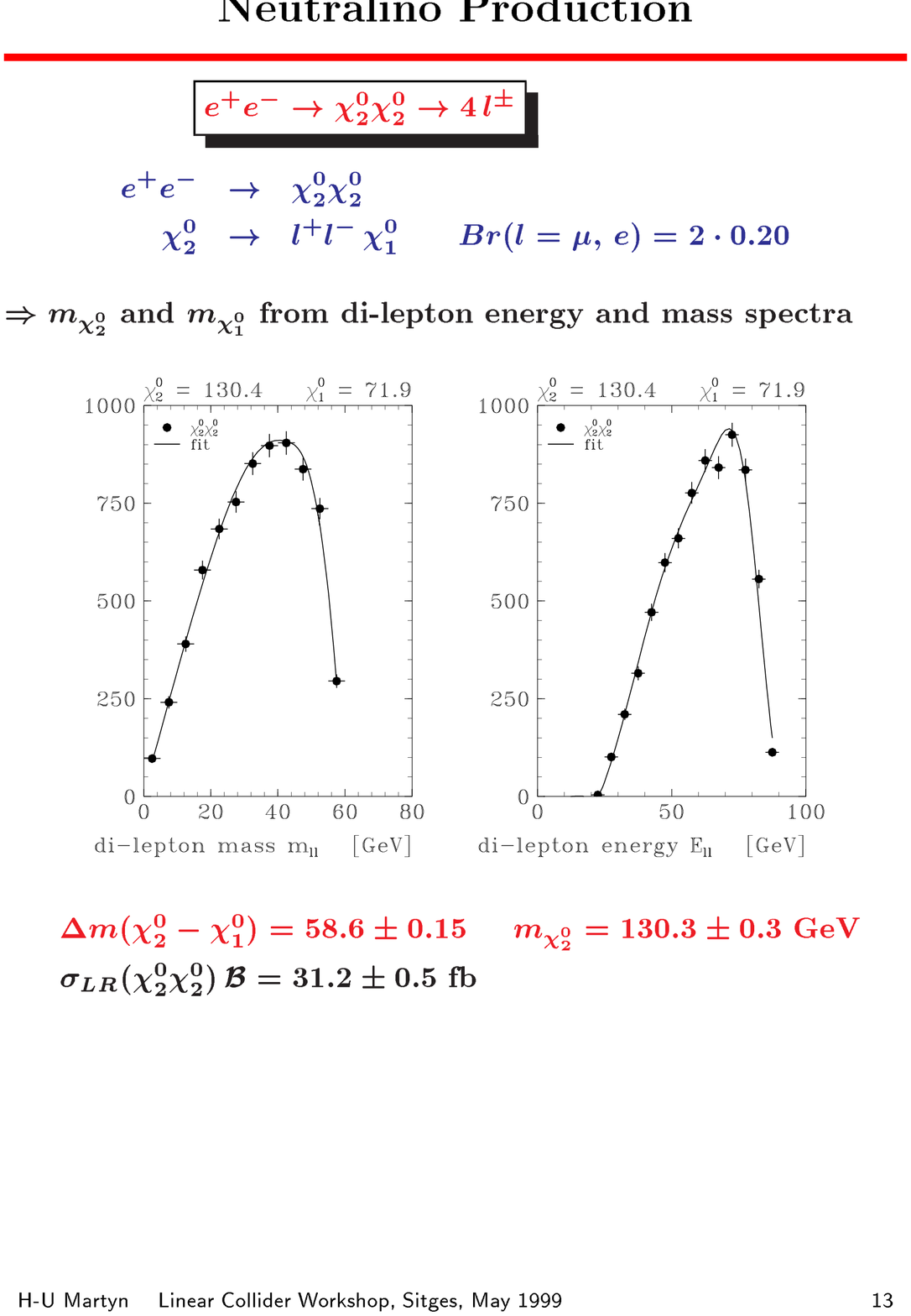,%
    bbllx=50pt,bblly=290pt,bburx=510pt,bbury=580pt,clip=,%
    angle=0,height=4cm}
  \hfill
  \epsfig{file=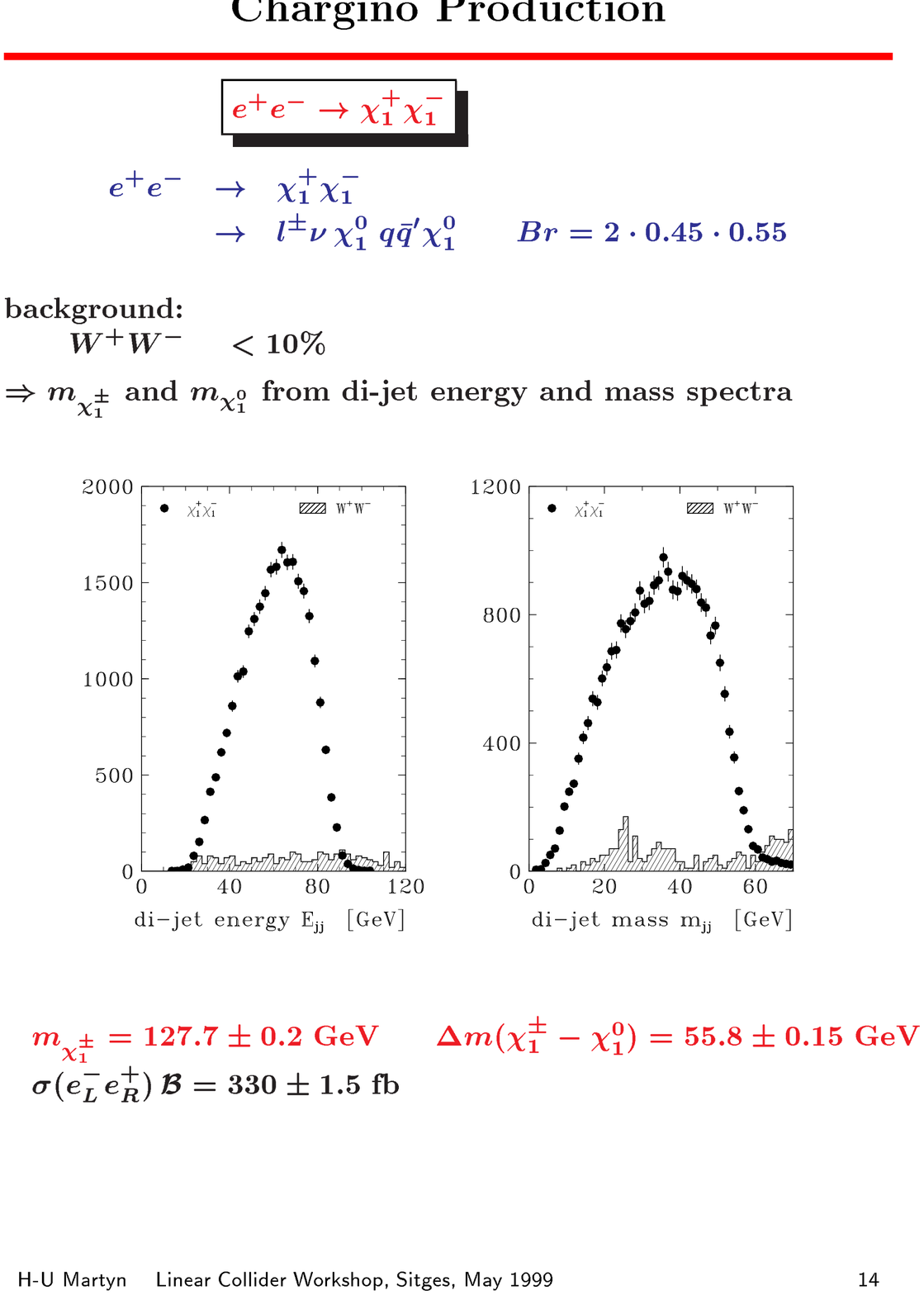,%
    bbllx=50pt,bblly=230pt,bburx=510pt,bbury=520pt,clip=,%
    angle=0,height=4cm} }
  \vspace{-.2cm}
  \caption{
    Di-lepton mass and energy spectra of
    $\chi^0_2 \to l^+ l^-\,\chi^0_1$ at $320~\GeV$ (left part) and
    di-jet mass and energy spectra of
    $\chi^\pm_1 \to q \bar{q}' \chi^0_1$ at $320~\GeV$ (right part).}
  \label{fig:gauginos}
\end{figure}

\noindent
{\bf Production of neutralinos and charginos} \quad
The lightest observable neutralino can be detected via its 3-body decay
$\chi^0_2 \rightarrow l^+ l^-\,\chi^0_1$. 
In a direct production the energy spectrum of the di-lepton system can be 
used to determine the masses of the primary and secondary neutralino,
similar to the slepton case. 
But $\chi^0_2$'s are also abundantly produced in decay chains and from
the upper edge of the di-lepton mass spectrum one gets a very precise
measurement of the mass difference 
$\Delta m (\chi^0_2 - \chi^0_1) = 58.6 \pm 0.05~\GeV$, 
essentially only limited by systematics.
Similarly, the copiously produced charginos with decays
$\chi^\pm_1 \to  q \bar{q}'\chi^0_1$ give
$\Delta m (\chi^\pm_1 - \chi^0_1) = 55.8 \pm 0.05~\GeV$.
The lowest mass neutralino production $e^-_R e^+_L \to \chi^0_2 \chi^0_1$ 
should be studied below $\tilde{l}_R \tilde{l}_R$ threshold, 
the polarisation suppresses $W^+W^-$ background.
The reaction $e^-_L e^+_R \to \chi^0_2 \chi^0_2 \to 4 l^\pm$ 
at $\sqrt{s}=320~\GeV$,
spectra shown in fig.~\ref{fig:gauginos}, 
is background free and gives a mass resolution of 
$\delta m_{\chi^0_2} = 0.3~\GeV$.
Another example is chargino production
$e^-_L e^+_R \to \chi^-_1 \chi^+_1 
             \to l^\pm \nu\,\chi^0_1 \ q \bar{q}'\chi^0_1$
with a small $W^+W^-$ contamination of $<10\%$,
see fig.~\ref{fig:gauginos}.
The large visible cross section, 
$\sigma_{LR} \,{\cal B} = 330$~fb,
compensates for the worse
jet energy resolution and results in an accuracy of
$\delta m_{\chi^\pm_1} = 0.2~\GeV$.
\begin{figure}[htb]
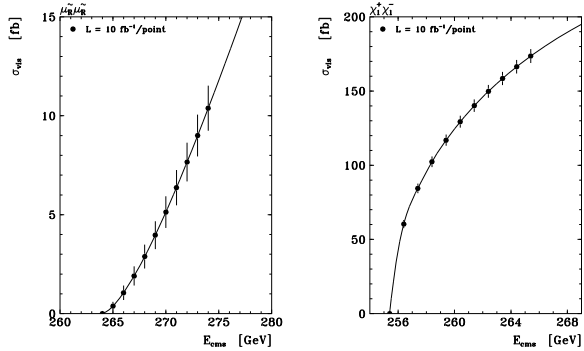

  \centering \vspace{-.3cm}
  \mbox{
  \epsfig{file=murscan.eps,%
    bbllx=0pt,bblly=400pt,bburx=500pt,bbury=800pt,clip=,%
    angle=90,height=5cm}
  \epsfig{file=c11scan.eps,%
    bbllx=0pt,bblly=400pt,bburx=500pt,bbury=800pt,clip=,%
    angle=90,height=5cm} }
  \vspace{-.2cm}
  \caption{ Visible cross sections near threshold of the reactions
    $e^-_R e^+_L \rightarrow \tilde{\mu}_R \tilde{\mu}_R$ (left) and
    $e^-_L e^+_R  \rightarrow \chi^-_1 \chi^+_1$ (right).
    Measurements assume ${\cal L} = 10~\fb^{-1}$ per point.}
  \label{fig:scan}
\end{figure}

\noindent
{\bf Threshold scans} \quad
Further improvement on sparticle masses may be achieved through 
threshold scans.
Such measurements are relatively simple, they essentially count additional
signatures over a smooth background.
Scans have been simulated by assuming a luminosity
of $100~\fb^{-1}$ distributed over 10 equidistant points, 
this procedure is not optimized.
The cross section for gaugino pair production rises $\propto \beta$, 
while the onset of slepton pair production is slower $\propto \beta^3$,
see excitation curves in fig.~\ref{fig:scan}.
Excellent mass resolutions below $100~\MeV$ can be obtained for 
the light chargino/neutralinos, degrading for the heavier $\chi$ states
to the per mil level.
Similar precisions can be obtained for the first generation sleptons
and right smuon. 
Higher mass states of the second and third generation suffer from low
detectable cross sections, but still accuracies of a few per mil 
are achievable.

The results of the various mass measurements using different techniques
are compiled in table~\ref{tab:masses}.
\begin{table}[htb]
  \caption{Expected precision on mass determinations using polarized
    $e^-$ and $e^+$ beams:
    $\delta m_{cont}$ from decay kinematics measured in the continuum
    (${\cal L}_{cont} = 160\,(250)~fb^{-1}$ 
    at $\sqrt{s} = 320\,(500)~\GeV$)
    and $\delta m_{scan}$ from threshold scans
    (${\cal L}_{scan} = 100~fb^{-1}$).
    The last column indicates the sensitivity to SUSY parameters.}
  \label{tab:masses}
  \vspace{.2cm}
    \begin{center}
      \begin{tabular}{l c c c l}
        \hdick \\[-1.5ex]
        particle
        & mass [GeV] & $\delta m_{cont}$ [GeV]  &  $\delta m_{scan}$  [GeV] 
        & SUSY parameters
        \\[1ex] \hdick  
        $\tilde{\mu}_R$   & 132.0 & 0.3 & 0.09 & 
        $\Rightarrow m_0, \ m_{1/2}, \ \phantom{\mu, \ }\tan\beta $ \\
        $\tilde{\mu}_L$   & 176.0 & 0.3 & 0.4 \\
        $\tilde{\nu}_\mu$ & 160.6 & 0.2 & 0.8 \\
        $\tilde{e}_R$     & 132.0 & 0.2 & 0.05 \\
        $\tilde{e}_L$     & 176.0 & 0.2 & 0.18 \\
        $\tilde{\nu}_e$   & 160.6 & 0.1 & 0.07 \\ 
        $\tilde{\tau}_1$  & 131.0 &     & 0.6 & 
        $\Rightarrow m_0, \ m_{1/2}, \ \mu, \ \tan\beta $ \\
        $\tilde{\tau}_2$  & 177.0 &     & 0.6 \\
        $\tilde{\nu}_\tau$& 160.6 &     & 0.6 \\ \hline
        $\chi^\pm_1$      & 127.7 & 0.2  & 0.04 & 
        $\Rightarrow  \phantom{M_1, \ }
                     M_2, \ \mu, \ \tan\beta $ \\
        $\chi^\pm_2$      & 345.8 &      & 0.25 \\ \hline
        $\chi^0_1$        &  71.9 & 0.1 & 0.05 & 
        $\Rightarrow M_1, \ M_2, \ \mu, \ \tan\beta $ \\
        $\chi^0_2$        & 130.3 & 0.3  & 0.07 \\
        $\chi^0_3$        & 319.8 &      & 0.30 \\
        $\chi^0_4$        & 348.2 &      & 0.52 
      \end{tabular}
    \end{center}
\end{table}

\section{Determination of SUSY parameters}

The precise mass measurements of sleptons, neutralinos and charginos
constitute an over-constrained set of observables which allow to determine
the structure and parameters of the underlying SUSY theory.
For example slepton universality can be readily tested to better 
than a percent.
The sensitivity of the particle masses to the MSSM or mSUGRA
parameters is indicated in table~\ref{tab:masses}.

In a first test the mSUGRA model is assumed, which
is based on the parameters
$m_0, \ m_{1/2}, \ A_0,\ \tan\beta$ and $\mbox{sign } \mu $.
The renormalisation group equations (RGE) are applied to extrapolate the 
masses to their common values at the GUT scale, 
where the  couplings of the gauge groups are related via
 $M_i/ \alpha_i =  m_{1/2}/ \alpha_{GUT}$.
Using the masses and their errors a $\chi^2$ fit is performed to determine
in a top-down approach the errors of the mSUGRA parameters assuming
the sign of $\mu$ to be known.
The magnitude of $\mu$ is obtained implicitly by the requirement
of electroweak symmetry breaking.
The common scalar and gaugino masses $m_0$ and $m_{1/2}$ can be determined
very precisely to better than a per mil, 
$\tan\beta$ to better than a percent and there is even some sensitivity
to the trilinear coupling $A_0$.
While much of the accuracy on $m_0$ amd $m_{1/2}$ can be obtained
using particle production which occur below thresholds of about 270~GeV for 
this model, the accuracy on $A_0$ and $\tan\beta$ requires using thresholds
up to about 400 GeV.

In a second fit the GUT relation between $M_1$ and $M_2$ is relaxed,
so that the same model is assumed under less constrained conditions.
As result $M_1$ and $M_2$ as well as $m_0$ can still be determined within
a per mil accuracy. 
The other parameters $\tan\beta$ and $A_0$ have slightly enlarged
uncertainty.

The results of both fits are summarized in the following table. 
It is obvious that the inclusion of cross section measurements will
improve these results.

\begin{center}
\begin{tabular}{l c c}
parameter &  true value & error\\
\hdick
{$m_0$}       & 100 GeV & {0.09 GeV}\\
{$m_{1/2}$}   & 200 GeV & {0.10 GeV}\\
& & \\
{$A_0$}       & 0 GeV   & {6.3 GeV}\\
{$\tan\beta$} & 3       & {0.02}\\
sgn($\mu$)    & +       & not fit\\
\end{tabular}
\hfill
\begin{tabular}{l c c}
parameter &  true value & error\\
\hdick
{$m_0$} & 100 GeV  & {0.09 GeV}\\
{$M_1$} & 200 GeV  & {0.20 GeV}\\
{$M_2$} & 200 GeV  & {0.20 GeV}\\
{$A_0$} & 0 GeV    & {10.3 GeV}\\
{$\tan\beta$} & 3  & {0.04}\\
sgn($\mu$) & + & not fit\\
\end{tabular}
\end{center}

\section{Conclusions}
The high luminosity of {\sc Tesla} allows to disentangle the production
and (cascade) decays of the accessible SUSY particle spectrum. 
Polarisation of both $e^-$ and $e^+$ beams is extremely important
for optimizing signal/background ratios.
Precision measurements of SUSY particle properties are achievable within 
reasonable time, e.g. masses with an accuracy of 
$\delta m \leq 0.3~\GeV$ in the continuum and
$\delta m \leq 0.1~\GeV$ at threshold.
Such measurements can be used in a RGE analysis of the mass spectrum to
determine the underlying SUSY model and parameters very accurately.

\end{document}